# Stable configuration of a molecule as spontaneous symmetry breaking

**V Damljanović**

Institute of Physics Belgrade, University of Belgrade, Pregrevica 118, 11080 Belgrade, Serbia

E-mail: damlja@ipb.ac.rs

**Abstract.** In a molecule subjected to no external fields, motion of nuclei is governed by a function $V$ of nuclear coordinates. This function (potential energy) is a sum of two terms: Coulomb repulsion between nuclei and the electronic effective potential $E$ which results from the Born-Oppenheimer approximation. In this paper we have presented a group of coordinate transformations which are the symmetries of functions $E$ and $V$. We showed that the formula for dynamical representation, which has fundamental importance in the symmetry analysis of normal modes of a molecule, follows from these symmetries. In addition, we proved that every molecule has at least one normal mode belonging to the totally symmetric (and therefore Raman-active) irreducible representation of the point group of that molecule. Next, we used symmetries of $V$ and $E$ to analyze possible shapes of some molecule types. We applied both Abud-Sartori-Michel theorem and symmetry adapted expansion of the electronic effective potential around united atom. Finally, we derived an approximate relation which predicts order of magnitude of the vibration frequency from the bond length in a diatomic molecule.

## 1. Introduction

Mathematical formalism in a spontaneous symmetry breaking theory relies on the existence of some function $f$ defined on a real m-dimensional vector space $\underline{R}^m$ and a compact group $G$ of unitary coordinate transformations that act in this vector space and are the symmetries of $f$. In other words:

$$(\forall \hat{g} \in G)(\forall \vec{x} \in \underline{R}^m) f(\hat{g}\vec{x}) = f(\vec{x}) \tag{1}$$

and we say that the function $f$ is $G$-invariant [1]. We are interested in the minima of $f$. Let $f$ reaches one of its minima at a point $\vec{x}_0$. From (1) it follows that this minimum is reached also at points from the set (called an orbit at $\vec{x}_0$) $\Omega(\vec{x}_0) \stackrel{def}{=} \{\hat{g}\vec{x}_0 | \hat{g} \in G\}$. The set of elements of $G$ which transforms $\vec{x}_0$ to itself will

Stable configuration of a molecule as spontaneous symmetry breaking be denoted $G_0$: $G_0 \stackrel{def}{=} \{\hat{g} \in G | \hat{g}\vec{x}_0 = \vec{x}_0\}$. $G_0$ is also a group (isotropy group of $\vec{x}_0$) and is a subgroup of $G$. For non-zero $\vec{x}_0$, $G_0$ is a true subgroup of $G$, we say that the minimum at $\vec{x}_0$ has the symmetry lower than $G$ and that the symmetry $G$ is spontaneously broken. In reality a physical system in which spontaneous symmetry breaking occurs chooses one element from the set $\Omega(\vec{x}_0)$ as a state of stable equilibrium although all other elements from this set are equally suitable.

In molecular physics one often applies Born-Oppenheimer approximation [2, 3] to separate the nuclear motion from the electronic. As a consequence of this approximation the Hamiltonian of nuclei is a sum of a kinetic energy term and an effective potential:

$$\hat{H}_n = \sum_{j=1}^{N} -\frac{\hbar^2}{2M_j}\Delta_j + V(\vec{R}_1, \vec{R}_2, ..., \vec{R}_N). \qquad (2)$$

In this formula $N$ is the number of nuclei in the molecule, $M_j$ is the mass of j-th nucleus and $\vec{R}_j$ is the position coordinate of the j-th nucleus. Stable equilibrium of the molecule is given by the minima of the function $V$. This function is a sum of Coulomb repulsion between nuclei and the electronic effective potential:

$$V(\vec{R}_1,...,\vec{R}_N) = \sum_{j=1}^{N-1}\sum_{l=j+1}^{N}\frac{q_j q_l}{4\pi\varepsilon_0|\vec{R}_j - \vec{R}_l|} + E(\vec{R}_1,...,\vec{R}_N), \qquad (3)$$

where $q_l$ is the charge of l-th nucleus. It turns out that $V$ and $E$ have the following symmetry properties:
a) Any two configurations (not necessary stable) of nuclei obtained from each other by rotation (proper and improper) of one configuration as a whole correspond to the same value of $V$ (respectively $E$).
b) Any two configurations (not necessary stable) of nuclei obtained from each other by a permutation between coordinates of identical nuclei correspond to the same value of $V$ (respectively $E$).
c) Any two configurations (not necessary stable) of nuclei obtained from each other by translation of one configuration as a whole correspond to the same value of $V$ (respectively $E$).

In this paper we will prove these three statements starting from the symmetries of a microscopic Hamiltonian. We will use this result to derive the formula for the dynamical representation. This representation was postulated by Eugene Wigner in 1930 with the justification that its action does not change relative distances between nuclei and is therefore the symmetry of the dynamical problem [4-6]. Here we will show that the formula for the dynamical representation can be obtained in another way. The advantage of our approach is that it will prove that for every molecule there exists at least one normal mode that belongs to the totally symmetric irreducible representation of the point group of that molecule. Another advantage is that the problem of finding minima of $V$ becomes another case of the spontaneous symmetry breaking phenomenon. This allows us to use the Abud-Sartori-Michel theorem [7-10] to discuss some stationary points of $V$. In the next part of the paper we will use symmetries we found to derive lowest order approximate expressions for the function $E$ and to discuss possible shapes of some molecular types that correspond to minima of $V$. Already in this simplest approximation one successfully obtains

Stable configuration of a molecule as spontaneous symmetry breaking some of molecular shapes that exist in nature. Our approach allows finding approximate relations between bond lengths in a molecule and its vibration frequencies. In the last part of the paper we derive this formula for the simplest case: that of a diatomic molecule.

**2. Symmetries of functions *E* and *V***
Our starting point is the Schroedinger equation for electrons moving in the field generated by nuclei in a molecule subjected to no external fields:

$$\left[\sum_{j=1}^{N_e} -\frac{\hbar^2}{2m_e}\cdot\left(\frac{\partial}{\partial \vec{r}_j^a}\right)^2 + \sum_{j=1}^{N_e}\sum_{l=1}^{N}\frac{eq_l}{4\pi\varepsilon_0\left|\vec{r}_j^a - \vec{R}_l^a\right|} + \sum_{j=1}^{N_e-1}\sum_{l=j+1}^{N_e}\frac{e^2}{4\pi\varepsilon_0\left|\vec{r}_j^a - \vec{r}_l^a\right|}\right]\psi(\vec{r}^a,\vec{R}^a) = E\psi(\vec{r}^a,\vec{R}^a). \quad (4)$$

In this equation $m_e$ is the electron mass, $N_e$ is the total number of electrons (the molecule need not be electrically neutral), $\vec{r}^a \stackrel{def}{=} (\vec{r}_1^a,...,\vec{r}_{N_e}^a)^T$ (the set of position coordinates of $N_e$ electrons, $T$ is transposition) is a vector in the $3N_e$-dimensional real vector space and $\vec{R}^a \stackrel{def}{=} (\vec{R}_1^a,...,\vec{R}_N^a)^T$ (the set of position coordinates of $N$ nuclei) is a vector in the $3N$-dimesional real vector space. The first term in the left hand side of (4) is kinetic energy of electrons, the second term is Coulomb attraction between electrons and nuclei and the third term is Coulomb repulsion between electrons. In the equation (4) coordinates of nuclei are parameters on which the eigenvalue $E$ is dependent. We write:

$$E = E(\vec{R}^a). \quad (5)$$

Let us now change variables in (4) in the following way:

$$\begin{aligned}(\forall j \in \{1,...,N_e\})\vec{r}_j^a = \hat{h}\vec{r}_j \\ (\forall l \in \{1,...,N\})\vec{R}_l^a = \hat{h}\vec{R}_l\end{aligned} \quad (6)$$

where $\hat{h}$ is an element from the orthogonal group: $\hat{h} \in \underline{O}(3)$. This is the group of all real, orthogonal, three-dimensional matrices. Due to orthogonality of $\hat{h}$ we have:

$$\left(\frac{\partial}{\partial \vec{r}_j^a}\right)^2 = \left(\frac{\partial}{\partial \vec{r}_j^a},\frac{\partial}{\partial \vec{r}_j^a}\right) = \left(\frac{\partial}{\partial(\hat{h}\vec{r}_j)},\frac{\partial}{\partial(\hat{h}\vec{r}_j)}\right) = \left(\hat{h}\frac{\partial}{\partial \vec{r}_j},\hat{h}\frac{\partial}{\partial \vec{r}_j}\right) = \left(\hat{h}^T\hat{h}\frac{\partial}{\partial \vec{r}_j},\frac{\partial}{\partial \vec{r}_j}\right) = \left(\frac{\partial}{\partial \vec{r}_j},\frac{\partial}{\partial \vec{r}_j}\right) = \left(\frac{\partial}{\partial \vec{r}_j}\right)^2 \quad (7)$$

where ( , ) denotes scalar product. For the same reason we have:

$$\left|\vec{r}_j^a - \vec{R}_l^a\right| = \left|\hat{h}(\vec{r}_j - \vec{R}_l)\right| = \left|\vec{r}_j - \vec{R}_l\right| \text{ and } \left|\vec{r}_j^a - \vec{r}_l^a\right| = \left|\hat{h}(\vec{r}_j - \vec{r}_l)\right| = \left|\vec{r}_j - \vec{r}_l\right| \quad (8)$$

and we use the fact that $\hat{h}$ is norm-conserving transformation. The equations (7) and (8) are valid for every $j$ and $l$. In a shorter notation the transformation (6) can be written in the following way:

$$\begin{array}{l} \vec{r}^a = \hat{w}\vec{r} \\ \vec{R}^a = \hat{g}_1 \vec{R} \end{array} \text{ with } \begin{array}{l} \hat{w} \overset{def}{=} \hat{I}_{N_e} \otimes \hat{h} \\ \hat{g}_1 \overset{def}{=} \hat{I}_N \otimes \hat{h} \end{array}. \quad (9)$$

In (9) $\hat{I}_N$ (respectively $\hat{I}_{N_e}$) is the $N$-dimensional (respectively $N_e$-dimensional) unit matrix and $\otimes$ denotes direct product. In the new variables the equation (4) reads:

$$\left[ \sum_{j=1}^{N_e} -\frac{\hbar^2}{2m_e} \cdot \left(\frac{\partial}{\partial \vec{r}_j}\right)^2 + \sum_{j=1}^{N_e} \sum_{l=1}^{N} \frac{eq_l}{4\pi\varepsilon_0 |\vec{r}_j - \vec{R}_l|} + \sum_{j=1}^{N_e-1} \sum_{l=j+1}^{N_e} \frac{e^2}{4\pi\varepsilon_0 |\vec{r}_j - \vec{r}_l|} \right] \psi(\hat{w}\vec{r}, \hat{g}_1\vec{R}) = E\psi(\hat{w}\vec{r}, \hat{g}_1\vec{R}). \quad (10)$$

Since transformed Hamiltonian looks the same as the original it is invariant under the substitution (6). In (10) new nuclear coordinates are also parameters so we can write:

$$E = E(\vec{R}). \quad (11)$$

Due to invariance of the Hamiltonian, functional dependence of the eigenvalue $E$ on $\vec{R}$ in (11) is the same as functional dependence of the eigenvalue $E$ on $\vec{R}^a$ in (5). From the equations (5) and (11) we have: $E(\vec{R}^a) = E(\vec{R})$, or:

$$\left(\forall \hat{h} \in O(3)\right)\left(\hat{g}_1 \overset{def}{=} \hat{I}_N \otimes \hat{h}\right)\left(\forall \vec{R} \in \underline{R}^{3N}\right) E(\hat{g}_1 \vec{R}) = E(\vec{R}). \quad (12)$$

We proved the statement a) from the introduction. By completely analogous reasoning one can prove statements b) and c). For the proof of the statement b) we will assume that in a molecule there are $s$ different chemical types of nuclei: $p_1$ nuclei of type one, $p_2$ nuclei of type 2,..., $p_s$ nuclei of type $s$. The sum of all $p$'s equals the number $N$ of all nuclei in the molecule:

$$p_1 + p_2 + ... + p_s = N. \quad (13)$$

Without loss of generality we will arrange nuclear coordinates in the following way: first $p_1$ coordinates are coordinates of type one nuclei, next $p_2$ coordinates are coordinates of type two nuclei and so on. The last $p_s$ coordinates are coordinates of type $s$ nuclei. With this assumption permutation between identical nuclei can be represented in matrix form in the following way:

Stable configuration of a molecule as spontaneous symmetry breaking

$$\vec{R}^a = \hat{g}_2 \vec{R} \text{ with } \hat{g}_2 \stackrel{def}{=} \hat{P}_N \otimes \hat{I}_3 \text{ and } \hat{P}_N \stackrel{def}{=} diag(\hat{S}_1(\pi_1), \hat{S}_2(\pi_2),...,\hat{S}_s(\pi_s)). \quad (14)$$

In (14) $\hat{S}_j(\pi_j)$ is the matrix of the permutation representation of the symmetric group $S_{p_j}$ that corresponds to the permutation $\pi_j \in S_{p_j}$. For example let $p_j=3$, $\pi_j = \begin{pmatrix} 123 \\ 231 \end{pmatrix}$ then permutation representation matrix that corresponds to this permutation is $\begin{pmatrix} 0 & 1 & 0 \\ 0 & 0 & 1 \\ 1 & 0 & 0 \end{pmatrix}^T$ because: $\begin{pmatrix} 2 \\ 3 \\ 1 \end{pmatrix} = \begin{pmatrix} 0 & 1 & 0 \\ 0 & 0 & 1 \\ 1 & 0 & 0 \end{pmatrix} \begin{pmatrix} 1 \\ 2 \\ 3 \end{pmatrix}$.

The $N$-dimensional matrix $\hat{P}_N$ has block-diagonal form with matrices of the permutation representation of corresponding symmetric groups on the main diagonal. $\hat{I}_3$ is the three-dimensional unit matrix. Since the Hamiltonian in (4) is invariant under the transformation (14) we have:

$$(\forall \pi_1 \in S_{p_1})..(\forall \pi_s \in S_{p_s}) \left( \hat{P}_N \stackrel{def}{=} diag(\hat{S}_1(\pi_1),...,\hat{S}_s(\pi_s)) \right) \left( \hat{g}_2 \stackrel{def}{=} \hat{P}_N \otimes \hat{I}_3 \right) (\forall \vec{R} \in \underline{R}^{3N}) E(\hat{g}_2 \vec{R}) = E(\vec{R}). \quad (15)$$

We proved the statement b) from the introduction. From (12) and (15) it follows that the general element of the symmetry group of $E$ is the product of $g_1$ and $g_2$: $\hat{g} \stackrel{def}{=} \hat{g}_1 \cdot \hat{g}_2 = (\hat{I}_N \otimes \hat{h}) \cdot (\hat{P}_N \otimes \hat{I}_3) = \hat{P}_N \otimes \hat{h}$ so that the following equation is valid:

$$(\forall \pi_1 \in S_{p_1})..(\forall \pi_s \in S_{p_s}) \left( \hat{P}_N \stackrel{def}{=} diag(\hat{S}_1(\pi_1),...,\hat{S}_s(\pi_s)) \right) (\forall \hat{h} \in \underline{O}(3))$$
$$\left( \hat{g} \stackrel{def}{=} \hat{P}_N \otimes \hat{h} \right) (\forall \vec{R} \in \underline{R}^{3N}) E(\hat{g}\vec{R}) = E(\vec{R}). \quad (16)$$

The total symmetry group of $E$ is the product: $G = S_{p_1} \times ... \times S_{p_s} \times \underline{O}(3)$ or more precisely the set $G$ of orthogonal matrices of the following form:

$$G = \left\{ \hat{g} \stackrel{def}{=} \hat{P}_N \otimes \hat{h} \middle| (\hat{h} \in \underline{O}(3)) \left( \hat{P}_N \stackrel{def}{=} diag(\hat{S}_1(\pi_1),...,\hat{S}_s(\pi_s)) \right) (\pi_1 \in S_{p_1})...(\pi_s \in S_{p_s}) \right\}. \quad (17)$$

Since Coulomb repulsion between nuclei in (3) has the same symmetry properties as $E$, the potential $V$ has also symmetry (16):

Stable configuration of a molecule as spontaneous symmetry breaking

$$\left(\forall \pi_1 \in S_{p_1}\right)..\left(\forall \pi_s \in S_{p_s}\right)\left(\hat{P}_N \stackrel{def}{=} diag\left(\hat{S}_1(\pi_1),...,\hat{S}_s(\pi_s)\right)\right)\left(\forall \hat{h} \in \underline{O}(3)\right)$$
$$\left(\hat{g} \stackrel{def}{=} \hat{P}_N \otimes \hat{h}\right)\left(\forall \vec{R} \in \underline{R}^{3N}\right) V(\hat{g}\vec{R}) = V(\vec{R}). \tag{18}$$

We have found the compact group $G$ of orthogonal transformations such that $E$ and $V$ are $G$-invariant. We will now use (18) to show the origin of the formula for the dynamical representation. The theory of molecular normal modes of oscillations relies on the Taylor expansion of the function $V$ around minimum up to terms of the second order. Let $\vec{R}^0$ be stable configuration of the molecule, i.e. the minimum of $V$ and $\vec{u} \stackrel{def}{=} (\vec{u}_1,...,\vec{u}_N)^T$ are small deviations from the stable configuration. We will define function $\overline{V}(\vec{u})$ in the following way:

$$\left(\forall \vec{u} \in \underline{R}^{3N}\right) \overline{V}(\vec{u}) \stackrel{def}{=} V(\vec{R}^0 + \vec{u}). \tag{19}$$

For $G_0$ being the isotropy group of $\vec{R}^0$:

$$G_0 \stackrel{def}{=} \left\{\hat{g} \in G \middle| \hat{g}\vec{R}^0 = \vec{R}^0\right\} \tag{20}$$

the function $\overline{V}$ has the following symmetry:

$$\left(\forall \hat{g} \in G_0\right)\left(\forall \vec{u} \in \underline{R}^{3N}\right) \overline{V}(\hat{g}\vec{u}) = V(\vec{R}^0 + \hat{g}\vec{u}) = V(\hat{g}(\vec{R}^0 + \vec{u})) = V(\vec{R}^0 + \vec{u}) = \overline{V}(\vec{u}). \tag{21}$$

The formula (21) is valid for all $\vec{u}$ (not necessary small ones) and we used the formulae (18) and (20). The elements of $G_0$ commute with the Hessian of $\overline{V}$ calculated at $\vec{u} = 0$:

$$\hat{D} \stackrel{def}{=} \left[\left|\frac{\partial}{\partial \vec{u}}\right\rangle\!\left\langle\frac{\partial}{\partial \vec{u}}\right| \overline{V}(\vec{u})\right]_{\vec{u}=0} = \left[\left|\frac{\partial}{\partial (\hat{g}\vec{u}^a)}\right\rangle\!\left\langle\frac{\partial}{\partial (\hat{g}\vec{u}^a)}\right| \overline{V}(\hat{g}\vec{u}^a)\right]_{\vec{u}^a=0}$$
$$= \left[\hat{g}\left|\frac{\partial}{\partial \vec{u}^a}\right\rangle\!\left\langle\frac{\partial}{\partial \vec{u}^a}\right| \hat{g}^T \overline{V}(\vec{u}^a)\right]_{\vec{u}^a=0} = \hat{g} \cdot \hat{D} \cdot \hat{g}^T = \hat{g} \cdot \hat{D} \cdot \hat{g}^{-1}. \tag{22}$$

For this reason elements of $G_0$ are symmetries of the dynamical problem of the molecule. In (22) we used the substitution $\vec{u} = \hat{g}\vec{u}^a$ and orthogonality of the elements of $G_0$.

From (17) and (20) it follows:

Stable configuration of a molecule as spontaneous symmetry breaking

$$\hat{g}\vec{R}^0 = (\hat{P}_N \otimes \hat{I}_3) \cdot (\hat{I}_N \otimes \hat{h}) \begin{pmatrix} \vec{R}_1^0 \\ . \\ . \\ . \\ \vec{R}_N^0 \end{pmatrix} = (\hat{P}_N \otimes \hat{I}_3) \begin{pmatrix} \hat{h}\vec{R}_1^0 \\ . \\ . \\ . \\ \hat{h}\vec{R}_N^0 \end{pmatrix} = \begin{pmatrix} \vec{R}_1^0 \\ . \\ . \\ . \\ \vec{R}_N^0 \end{pmatrix}. \tag{23}$$

Since the matrix $\hat{P}_N$ only permutes the components of the vector $\vec{R}^0$ the equation (23) will have solution if and only if an element $\hat{h}$ from $\underline{O}(3)$ is such that for every coordinate $\vec{R}_j^0$ from the set $\{\vec{R}_1^0,...,\vec{R}_N^0\}$, $\hat{h}\vec{R}_j^0$ belongs to the same set and to the nucleus of the same type. Then we have to pick up an element from the symmetric group $S_{p_1} \times ... \times S_{p_s}$ that will put coordinates back in the original order. Such an element will depend on $\hat{h}$ so finally we have:

$$(\forall \hat{h} \in H \subset \underline{O}(3))[\hat{P}_N(\hat{h}) \otimes \hat{h}] \begin{pmatrix} \vec{R}_1^0 \\ . \\ . \\ . \\ \vec{R}_N^0 \end{pmatrix} = \begin{pmatrix} \vec{R}_1^0 \\ . \\ . \\ . \\ \vec{R}_N^0 \end{pmatrix}. \tag{24}$$

$H$ is the group of all elements of $\underline{O}(3)$ for which (23) has a solution. Matrix $\hat{P}_N(\hat{h}) \otimes \hat{h}$ in (24) is exactly dynamical representations of the group $H$ with $H$ being the point group of the molecule. In addition, from (24) follows that the stable configuration $\vec{R}^0$ belongs to the totally symmetric irreducible representation of $H$ i.e. to the representation in which all elements of $H$ are represented by the number one. For this reason (24) will have the only solution $\vec{R}^0 = 0$ if the dynamical representation does not contain the totally symmetric irreducible representation. Since this is non-physical situation it follows that in every molecule, the dynamical representation of the point group of that molecule contains at least one totally symmetric irreducible representation. In other words for every molecule there exist at least one normal mode that during motion conserves the symmetry of the stable configuration of that molecule.

Finally let's prove the statement c) from the introduction. This statement follows from the invariance of the Hamiltonian in (4) under transformation:

$$\begin{aligned}(\forall j \in \{1,...,N_e\})\vec{r}_j^a &= \vec{r}_j + \vec{t} \\ (\forall l \in \{1,...,N\})\vec{R}_l^a &= \vec{R}_l + \vec{t}\end{aligned} \tag{25}$$

$\vec{t}$ is any element from a real, three dimensional vector space $\underline{R}^3$. By applying the same train of thought as in derivation of the formula (12) we get the following equality:

$$(\forall \vec{t} \in \underline{R}^3)\left(\vec{\tau} \stackrel{def}{=} (\vec{t},\vec{t},...,\vec{t})^T\right)(\forall \vec{R} \in \underline{R}^{3N})E(\vec{R}+\vec{\tau}) = E(\vec{R}). \tag{26}$$

Stable configuration of a molecule as spontaneous symmetry breaking

Since the same symmetry property holds for the Coulomb repulsion between nuclei in (3), it holds for the function $V$ too:

$$\left(\forall \vec{t} \in \underline{R}^3\right)\left(\vec{\tau} \stackrel{def}{=} (\vec{t},\vec{t},...,\vec{t})^T\right)\left(\forall \vec{R} \in \underline{R}^{3N}\right) V(\vec{R}+\vec{\tau}) = V(\vec{R}) . \tag{27}$$

This concludes derivation of symmetry properties of $V$ and $E$. The equation (18) allows us to use formalism developed by Abud and Sartori [7-9] for finding sufficient conditions for a vector from $\underline{R}^{3N}$ to be a stationary point of $G$-invariant function $V$. The next section is devoted to this question.

### 3. Stationary points of $V$ from Abud-Sartori-Michel theorem

We will use Abud-Sartori formulation of Michel's theorem [7-9]. Stationary points of $V$ are points for which gradient of $V$ is equal to zero. Abud-Sartori formulation of Michel's theorem gives sufficient (but not necessary) conditions for a $G$-invariant function to have a stationary point. Two subgroups of $G$, $G_1$ and $G_2$ are conjugated in $G$ if there exist an element $g$ from $G$ such that $G_1 = gG_2g^{-1}$. The isotropy subgroups at points lying on the same orbit are conjugated in $G$. Therefore it is meaningful to assign isotropy group to an orbit. Two orbits belong to the same orbit type if (by definition) their isotropy groups are conjugated in $G$. The union of all orbits having the same orbit type is called a stratum of $\underline{R}^m/G$. Since $G$ is a compact group acting orthogonally on a real m-dimensional vector space there exist a minimal set of $G$-invariant polynomials $\theta_1,...,\theta_k$ such that any other invariant polynomial is a polynomial over $\theta_1,...,\theta_k$. This set is called minimal integrity basis (MIB). MIB maps different orbits into different points of $D \subset \underline{R}^k$. Using MIB one defines the following important relation of partial ordering between strata: stratum $\Sigma_2$ will be said to be more peripheral than $\Sigma_1$ if some connected component of $\Sigma_2$ is contained in the closure of $\Sigma_1$. A stratum $\Sigma$ will be maximally peripheral if in its closure there are no connected components of other strata except trivial. Trivial stratum consists of zero vector only and has the whole $G$ as the isotropy group. Abud-Sartori formulation of Michel's theorem states that, under some general conditions, $G$-invariant function $V$ will have stationary point on every connected component of every maximally peripheral stratum [9]. We apply this theorem to few examples.

*A molecule of the type XY.* For this type of molecule the symmetry group is

$$G = \left\{ \begin{pmatrix} \hat{h} & \hat{0} \\ \hat{0} & \hat{h} \end{pmatrix} \middle| \hat{h} \in \underline{O}(3) \right\}, \tag{28}$$

acting orthogonally on elements

$$\vec{R} = \begin{pmatrix} \vec{R}_1 \\ \vec{R}_2 \end{pmatrix} \tag{29}$$

Stable configuration of a molecule as spontaneous symmetry breaking

of 6-dimensional real vector space. MIB consists of the following 3 polynomials:

$$\theta_1 = \vec{R}_1^2, \quad \theta_2 = \vec{R}_2^2 \text{ and } \theta_3 = \vec{R}_1 \circ \vec{R}_2. \tag{30}$$

The range $D$ from the space $\underline{R}^3$ into which $\underline{R}^6$ is mapped by the MIB is given by the following inequalities:

$$0 \leq \theta_1 < +\infty, \quad 0 \leq \theta_2 < +\infty, \quad -\sqrt{\theta_1 \theta_2} \leq \theta_3 \leq \sqrt{\theta_1 \theta_2}, \tag{31}$$

where the last expression follows from the Schwartz inequality. For every two non negative real values $a$ and $b$ and for every $\alpha$ between zero and $\pi$, an orbit is defined in the following way:

$$\Omega_{a,b,\alpha} = \left\{ \begin{pmatrix} \vec{R}_1 \\ \vec{R}_2 \end{pmatrix} \middle| \left|\vec{R}_1\right| = a, \left|\vec{R}_2\right| = b, \angle(\vec{R}_1, \vec{R}_2) = \alpha \right\}. \tag{32}$$

There are three strata:

$$\Sigma_1 = \{\Omega_{a,b,\alpha} | a > 0, b > 0, \pi > \alpha > 0\} \text{ with symmetry } \underline{C}_s,$$
$$\Sigma_2 = \{\Omega_{a,b,\alpha} | a \geq 0, b \geq 0, a+b > 0, \alpha = 0 \vee \alpha = \pi\} \text{ with symmetry } \underline{C}_{\infty v} \text{ and}$$
$$\Sigma_3 = \{\Omega_{a,b,\alpha} | a = b = 0\} \text{ with the symmetry } \underline{O}(3).$$

$\Sigma_2$ is in closure of $\Sigma_1$ because $\alpha=0$ and $\alpha=\pi$ are in the closure of the interval $(0, \pi)$. $\Sigma_3$ (the trivial stratum) is the only stratum in closure of $\Sigma_2$. For this reason $\Sigma_2$ is maximally peripheral stratum and $V$ will have a stationary point on this stratum with symmetry $\underline{C}_{\infty v}$.

*A molecule of the type $\underline{X}_2$.* The symmetry group is

$$G = \left\{ \begin{pmatrix} \hat{h} & \hat{0} \\ \hat{0} & \hat{h} \end{pmatrix}, \begin{pmatrix} \hat{0} & \hat{h} \\ \hat{h} & \hat{0} \end{pmatrix} \middle| \hat{h} \in \underline{O}(3) \right\}, \tag{33}$$

acting orthogonally on elements

$$\vec{R} = \begin{pmatrix} \vec{R}_1 \\ \vec{R}_2 \end{pmatrix} \tag{34}$$

of 6-dimensional real vector space. MIB has 3 polynomials:

$$\theta_1 = \vec{R}_1^2 + \vec{R}_2^2, \quad \theta_2 = \vec{R}_1 \circ \vec{R}_2 \text{ and } \theta_3 = \vec{R}_1^2 \cdot \vec{R}_2^2. \tag{35}$$

Stable configuration of a molecule as spontaneous symmetry breaking

We can proceed with our analysis even without knowing MIB. We have to build configurations of different symmetries from points $\vec{R}_1$, $\vec{R}_2$ and the origin and to see which configurations are maximally peripheral. The stratum $\Sigma_1$ contains configurations in which the origin and the two *X* nuclei are located on a plane such that this plane is the only symmetry. The symmetry of such configuration is $C_s$. From this configuration we can by deformation arrive at following two strata: $\Sigma_2$ (nuclei are at the same distance from the origin but the origin and the nuclei are not on the same axis) with symmetry $C_{2v}$ and $\Sigma_3$ (nuclei and the origin are on the same axis but no other symmetries exist except the axis and planes that contains that axis) with symmetry $C_{\infty v}$. Both $\Sigma_2$ and $\Sigma_3$ are more peripheral than $\Sigma_1$. From these two strata we arrive by deformation to stratum $\Sigma_4$ (the origin is located in the middle between two nuclei) with the symmetry $D_{\infty h}$. It follows that $\Sigma_4$ is more peripheral than $\Sigma_2$ and $\Sigma_3$. Since the trivial stratum is the only stratum in the closure of $\Sigma_4$ it follows that the $\Sigma_4$ is the maximally peripheral stratum. *V* will therefore have stationary point on this stratum with symmetry $D_{\infty h}$.

*A molecule of the type XY$_2$*. The strata are of the following symmetries: $C_1$ (nuclei and the origin do not belong to the same plane), $C_s$ (nuclei and the origin do belong to the same plane but no other symmetries exist), $C_{2v}$ (the nucleus *X* is at the origin and *Y* nuclei are at the same distance from the origin but all nuclei do not belong to the same axis), $C_{\infty v}$ (all three nuclei and the origin are located on the same axis but in such a way that the axis and planes that contain it are the only symmetries) and $D_{\infty h}$ (the nucleus *X* is at the origin and in the middle between two *Y* nuclei, all three are located on the same axis). The last stratum is more peripheral than $C_{\infty v}$ and $C_{2v}$, these are more peripheral than $C_s$ and $C_s$ is more peripheral than $C_1$. It follows that $D_{\infty h}$ is maximally peripheral stratum, so *V* will have stationary point on it. As an example of a molecule of the type *XY$_2$* with symmetry $D_{\infty h}$ is $CO_2$.

*A molecule of the type X$_3$*. Maximally peripheral strata are $D_{\infty h}$ (all three nuclei belong to the same axis, one is at the origin and in the middle between the remaining two) and $D_{3h}$ (three nuclei are at vertices of equilateral triangle with its center at the origin). As an example of molecule of the type *X$_3$* with the symmetry $D_{\infty h}$ is $C_3$, and with the symmetry $D_{3h}$ the molecule $H_3^+$.

*A molecule of the type XY$_3$*. Maximally peripheral stratum is $D_{3h}$ (the *X* nucleus is at the origin and *Y*-nuclei are at vertices of equilateral triangle with its center at the origin). An example of such configuration in nature is the molecule $BH_3$.

*A molecule of the type X$_4$*. There are 4 maximally peripheral strata. $D_{\infty h}$ (all 4 nuclei and the origin, are located on an axis, nuclei are symmetrically positioned around the origin). An example is the molecule $C_4$. $T_d$ (all 4 nuclei are at vertices of regular tetrahedron with its center at the origin), with the molecule $P_4$ as an example. $D_{4h}$ (4 nuclei at vertices of a square, with its center at origin) and $D_{3h}$ (one nucleus at the origin, remaining 3 at vertices of equilateral triangle. The center of the triangle is at the origin). For the last two strata there are no examples in nature, therefore they most likely correspond either to the maximum of *V* or to its saddle point.

We have considered only configurations in which all nuclear coordinates are different to each other, since the cases where two or more nuclei are located at the same point are non physical. Since translation

Stable configuration of a molecule as spontaneous symmetry breaking is a non-linear operation its action on a vector cannot be written in a form of a matrix acting on that vector. Because of that we couldn't use symmetry given by (27) in discussing stationary points of *V* from the point of view of Abud-Sartori-Michel theorem. In addition, this theorem does not give all stationary points of *V* and it cannot predict which stationary points correspond to a minimum. For that reason it is important to know more details about the function *V*. In the next section we will use symmetry adapted approximate expressions of the lowest order for the function *E* to find minima of *V*. Symmetry relations (16) and (26) will significantly reduce the number of unknown coefficients in the expansion of *E* with respect to nuclear coordinates.

## 4. Minima of *V* by use of approximate expressions for *E*

Since Coulomb repulsion between nuclei diverges at origin, the function *V* cannot be expanded by use of Taylor formula around that point. On the other hand, *E* is finite at the point $\vec{R}=0$. It is equal to the electronic energy of an atom (called united atom) having charge of the nucleus equal to sum of charges of all nuclei in the molecule under investigation. Moreover, stable configurations of relatively simple molecules are confined in space in the small volume around an origin (center of mass of the molecule). For these reasons it is possible to expand the function *E* around the origin with respect to nuclear coordinates. This expansion must contain only polynomials invariant under the action of the group *G* in (17) with additional symmetry given by the formula (26). In finding such polynomials one might use the following two facts:

1) Elementary symmetric polynomials defined in the following way

$$\sigma_1 \stackrel{def}{=} x_1 + x_2 + \ldots + x_n$$
$$\sigma_2 \stackrel{def}{=} x_1 x_2 + x_1 x_3 + \ldots + x_{n-1} x_n = \sum_{i<j} x_i x_j$$
$$\vdots$$
$$\sigma_n \stackrel{def}{=} x_1 x_2 \cdots x_n$$

form a complete system of invariants of the symmetric group $S_n$ [11].

2) For the group $\{\hat{I}_N \otimes \hat{h} | \hat{h} \in O(3)\}$ acting on vectors $\vec{R} = (\vec{R}_1,\ldots,\vec{R}_N)^T$ from a real 3*N*-dimensional vector space, the minimal integrity basis is $\{\vec{R}_j \circ \vec{R}_l | j = \overline{1,N}; l = \overline{1,j}\}$.

From the last statement follows that the expansion of *E* with respect to nuclear coordinates in every molecule will contain only polynomials of even degree. Here we will assume that the function *E* has derivatives at zero up to the second degree (this follows from [12-15]).

*Molecules of types XY and X₂*. For both types of molecules the function *V* is approximately:

$$V \approx a(\vec{R}_1 - \vec{R}_2)^2 + \frac{q_1 q_2}{4\pi\varepsilon_0 |\vec{R}_1 - \vec{R}_2|}, \tag{36}$$

Stable configuration of a molecule as spontaneous symmetry breaking where $a$ is a real parameter which we'll assume to be different than zero. For $a$ negative, the potential (36) does not have a minimum. For $a$ positive, the stable configuration is

$$\left|\vec{R}_1^0 - \vec{R}_2^0\right| = \left[\frac{q_1 q_2}{8\pi\varepsilon_0 a}\right]^{1/3}. \tag{37}$$

This configuration has $\underline{C}_{\infty v}$ symmetry for $\underline{XY}$ and $\underline{D}_{\infty h}$ for $\underline{X_2}$.

*A molecule of the type $\underline{XY_2}$.* Let the nucleus $\underline{X}$ has coordinate $\vec{R}_1$ and the charge $q$, nuclei $\underline{Y}$ have the charge $Q$ and coordinates $\vec{R}_2$ and $\vec{R}_3$. The function $V$ is approximately:

$$V \approx a_1\left[(\vec{R}_1 - \vec{R}_2)^2 + (\vec{R}_1 - \vec{R}_3)^2\right] + c_1(\vec{R}_2 - \vec{R}_3)^2 + \frac{1}{4\pi\varepsilon_0}\left[\frac{qQ}{\left|\vec{R}_1 - \vec{R}_2\right|} + \frac{qQ}{\left|\vec{R}_1 - \vec{R}_3\right|} + \frac{Q^2}{\left|\vec{R}_2 - \vec{R}_3\right|}\right]. \tag{38}$$

Here $a_1$ and $c_1$ are real parameters by assumption different than zero. In order to find minima of $V$ we will introduce the substitution:

$$\vec{\rho}_1 \stackrel{def}{=} \vec{R}_1 - \vec{R}_2$$
$$\vec{\rho}_2 \stackrel{def}{=} \vec{R}_1 - \vec{R}_3 \tag{39}$$
$$\vec{\rho}_3 \stackrel{def}{=} \vec{R}_1 + \vec{R}_2 + \vec{R}_3.$$

This substitution is invertible so its Jacobian is different than zero. For physical reasons $\vec{\rho}_1 \neq 0$, $\vec{\rho}_2 \neq 0$ and $\vec{\rho}_1 \neq \vec{\rho}_2$. In the new coordinates $V$ reads:

$$V \approx a_1(\vec{\rho}_1^2 + \vec{\rho}_2^2) + c_1(\vec{\rho}_1 - \vec{\rho}_2)^2 + \frac{1}{4\pi\varepsilon_0}\left[\frac{qQ}{|\vec{\rho}_1|} + \frac{qQ}{|\vec{\rho}_2|} + \frac{Q^2}{|\vec{\rho}_1 - \vec{\rho}_2|}\right]. \tag{40}$$

Let $\vec{\rho}_1^0$, $\vec{\rho}_2^0$ denote stable configuration. We will expand $V$ around this point up to the second order. For that purpose the following formula is useful:

$$\frac{1}{|\vec{\rho} + \delta\vec{\rho}|} \approx \frac{1}{|\vec{\rho}|} - \frac{\vec{\rho} \circ \delta\vec{\rho}}{|\vec{\rho}|^3} - \frac{1}{2}\cdot\frac{(\delta\vec{\rho})^2}{|\vec{\rho}|^3} + \frac{3}{2}\cdot\frac{(\vec{\rho} \circ \delta\vec{\rho})^2}{|\vec{\rho}|^5} \tag{41}$$

Stable configuration of a molecule as spontaneous symmetry breaking approximately valid for every non-zero $\vec{\rho}$ from the real 3-dimensional vector space and every small $\delta\vec{\rho}$. Since the stable configurations correspond to the minimum of $V$, terms linear in $\delta\vec{\rho}_1$ and $\delta\vec{\rho}_2$ are equal to zero:

$$\left[2a_1 + 2c_1 - \frac{qQ}{4\pi\varepsilon_0|\vec{\rho}_1^0|^3} - \frac{Q^2}{4\pi\varepsilon_0|\vec{\rho}_1^0 - \vec{\rho}_2^0|^3}\right] \cdot \vec{\rho}_1^0 + \left[-2c_1 + \frac{Q^2}{4\pi\varepsilon_0|\vec{\rho}_1^0 - \vec{\rho}_2^0|^3}\right] \cdot \vec{\rho}_2^0 = 0$$

$$\left[-2c_1 + \frac{Q^2}{4\pi\varepsilon_0|\vec{\rho}_1^0 - \vec{\rho}_2^0|^3}\right] \cdot \vec{\rho}_1^0 + \left[2a_1 + 2c_1 - \frac{qQ}{4\pi\varepsilon_0|\vec{\rho}_2^0|^3} - \frac{Q^2}{4\pi\varepsilon_0|\vec{\rho}_1^0 - \vec{\rho}_2^0|^3}\right] \cdot \vec{\rho}_2^0 = 0.$$

(42)

We can distinguish two cases:

The first case is for $\vec{\rho}_1^0$ and $\vec{\rho}_2^0$ non-collinear i.e. linearly independent. Then the only solution of equation (42) is that each coefficient next to $\vec{\rho}_1^0$ and $\vec{\rho}_2^0$ is equal to zero. That corresponds to the configuration with symmetry $C_{2v}$ in which the distance between $X$ and any of two $Y$ nuclei is equal to $\left[\dfrac{qQ}{8\pi\varepsilon_0 a_1}\right]^{1/3}$ while the distance between two $Y$ nuclei is $\left[\dfrac{Q^2}{8\pi\varepsilon_0 c_1}\right]^{1/3}$. This solution exists for $a_1$ and $c_1$ positive. For this case $V$ is approximately equal to:

$$V \approx V_0 + \frac{3}{8\pi\varepsilon_0}\left[\frac{qQ(\vec{\rho}_1^0 \circ \delta\vec{\rho}_1)^2}{|\vec{\rho}_1^0|^5} + \frac{qQ(\vec{\rho}_2^0 \circ \delta\vec{\rho}_2)^2}{|\vec{\rho}_2^0|^5} + \frac{Q^2[(\vec{\rho}_1^0 - \vec{\rho}_2^0) \circ (\delta\vec{\rho}_1 - \delta\vec{\rho}_2)]^2}{|\vec{\rho}_1^0 - \vec{\rho}_2^0|^5}\right].$$  (43)

Here $V_0$ is the value of $V$ calculated at the equilibrium. For every deformation of this configuration the second term in (43) is positive so this equilibrium corresponds to the minimum of $V$ and the configuration is stable. An example of a molecule of the type $XY_2$ with the symmetry $C_{2v}$ is the molecule $H_2O$.

The second case is for $\vec{\rho}_1^0$ and $\vec{\rho}_2^0$ collinear i.e. linearly dependent. It follows:

$$\vec{\rho}_2^0 = y\vec{\rho}_1^0,$$  (44)

where $y$ is a real number different from zero and one. It turns out that the only solution stable under certain restrictions on $a_1$ and $c_1$ is for $y$ equal to minus one. This case corresponds to the linear configuration with the symmetry $D_{\infty h}$ and the distance between $X$ and any of two $Y$ nuclei equal to:

$$\left[\frac{qQ + Q^2/4}{2a_1 + 4c_1}\right]^{1/3}.$$

Stable configuration of a molecule as spontaneous symmetry breaking

From this follows:

$$a_1 + 2c_1 > 0. \tag{45}$$

For the equilibrium in the second case $V$ is approximately equal to:

$$V \approx V_0 + \frac{1}{2}\frac{a_1 Q - 8c_1 q}{Q + 4q}(\delta\vec{\rho}_1 + \delta\vec{\rho}_2)^2$$
$$+ \frac{3}{8\pi\varepsilon_0 |\vec{\rho}_1^0|^5}\left[qQ(\vec{\rho}_1^0 \circ \delta\vec{\rho}_1)^2 + qQ(\vec{\rho}_1^0 \circ \delta\vec{\rho}_2)^2 + \frac{Q^2}{8}[\vec{\rho}_1^0 \circ (\delta\vec{\rho}_1 - \delta\vec{\rho}_2)]^2\right]. \tag{46}$$

The linear configuration will be stable if the second term in (46) is non-negative. This will be the case for:

$$a_1 Q - 8c_1 q > 0. \tag{47}$$

From (45) and (47) follows the criterion for the linear configuration to be stable:

$$\begin{aligned}&a_1 > 0\\&a_1 Q/(8q) > c_1 > -a_1/2.\end{aligned} \tag{48}$$

*A molecule of the type $X_3$*. For this molecule $V$ is approximately:

$$V \approx a_2\left[(\vec{R}_1 - \vec{R}_2)^2 + (\vec{R}_1 - \vec{R}_3)^2 + (\vec{R}_2 - \vec{R}_3)^2\right] + \frac{q^2}{4\pi\varepsilon_0}\left[\frac{1}{|\vec{R}_1 - \vec{R}_2|} + \frac{1}{|\vec{R}_1 - \vec{R}_3|} + \frac{1}{|\vec{R}_2 - \vec{R}_3|}\right]. \tag{49}$$

Here $a_2$ is by assumption non-zero real value. We can use the case of $XY_2$ to conclude that for positive $a_2$ stable configuration will be of the form of equilateral triangle with the distance between any two nuclei equal to $\left[\frac{q^2}{8\pi\varepsilon_0 a_2}\right]^{1/3}$. Corresponding symmetry is $D_{3h}$. The linear configuration will be unstable in the approximation we used, since (48) applied to the case of a molecule $X_3$ leads to a contradiction.

*A molecule of the type $XY_3$*. Let the nucleus $X$ has the charge $q$ and the coordinate $\vec{R}_1$, while nuclei $Y$ have the charge $Q$ and coordinates $\vec{R}_2$, $\vec{R}_3$ and $\vec{R}_4$. The lowest order approximation for $E$ gives for $V$:

$$V \approx a_3\left[(\vec{R}_1 - \vec{R}_2)^2 + (\vec{R}_1 - \vec{R}_3)^2 + (\vec{R}_1 - \vec{R}_4)^2\right] + c_3\left[(\vec{R}_2 - \vec{R}_3)^2 + (\vec{R}_2 - \vec{R}_4)^2 + (\vec{R}_3 - \vec{R}_4)^2\right]$$
$$+ \frac{1}{4\pi\varepsilon_0}\left[\frac{qQ}{|\vec{R}_1 - \vec{R}_2|} + \frac{qQ}{|\vec{R}_1 - \vec{R}_3|} + \frac{qQ}{|\vec{R}_1 - \vec{R}_4|} + \frac{Q^2}{|\vec{R}_2 - \vec{R}_3|} + \frac{Q^2}{|\vec{R}_2 - \vec{R}_4|} + \frac{Q^2}{|\vec{R}_3 - \vec{R}_4|}\right]. \tag{50}$$

Stable configuration of a molecule as spontaneous symmetry breaking

Here $a_3$, $c_3$ are by assumption non-zero real values. By a derivation analogous to the case of $\underline{XY_2}$ we conclude that there are two stable configurations. A pyramidal one for positive $a_3$, $c_3$, with the distance between the $\underline{X}$ and any of $\underline{Y}$ nuclei equal to $\left[\dfrac{qQ}{8\pi\varepsilon_0 a_3}\right]^{1/3}$, and the distance between any two $\underline{Y}$ nuclei equal to $\left[\dfrac{Q^2}{8\pi\varepsilon_0 c_3}\right]^{1/3}$. Such configuration has the symmetry $\underline{C_{3v}}$ with the molecule NH$_3$ as an example. Second configuration is planar with nuclei $\underline{Y}$ located at vertices of equilateral triangle and the nucleus $\underline{X}$ in its center. The distance between the nucleus $\underline{X}$ and any of $\underline{Y}$ nuclei is equal to $\left[\dfrac{qQ+Q^2/\sqrt{3}}{8\pi\varepsilon_0(a_3+3c_3)}\right]^{1/3}$. This is possible for

$$a_3 > 0$$
$$a_3 Q/(3\sqrt{3}q) > c_3 > -a_3/3. \qquad (51)$$

The symmetry of this configuration is $\underline{D_{3h}}$. An example is given in the section 3.

*A molecule of the type $\underline{X_4}$.* In our lowest order approximation for $E$ the only stable configuration will be that of the regular tetrahedron. Remaining configurations discussed in the section 3 are saddle points.

We have predicted some of molecular shapes that exist in nature already in the simplest approximation for the electronic effective potential $E$. For prediction of other molecular shapes (for example that of ozone O$_3$ which is of the type $\underline{X_3}$ but has the symmetry $\underline{C_{2v}}$) one has to go to higher other corrections for $E$.

Our lowest order expansion of $E$ with respect to coordinates of nuclei in a molecule allows approximate determination of its force-constants. In some cases, we need only to know bond lengths in a stable configuration of a molecule and the charges of nuclei. From the force-constants one obtains vibration frequencies in the usual way. In the next section we derive the connection between the equilibrium bond length and the only vibration frequency of a diatomic molecule.

**5. Approximate relation between the bond length and vibration frequency of a diatomic molecule**

Using (36) and the expansion (41) we get for the potential $V$ up to the second order terms around equilibrium configuration the following expression:

$$V \approx V_0 + \frac{3q_1 q_2 (\vec{\rho}_0 \circ \delta\vec{\rho})^2}{8\pi\varepsilon_0 |\vec{\rho}_0|^5}, \qquad (52)$$

where $\vec{\rho}_0 \stackrel{def}{=} \vec{R}_1^0 - \vec{R}_2^0$ and $|\vec{\rho}_0|$ is the equilibrium bond length. We will take the z-axis along the axis of the molecule i.e. along $\vec{\rho}_0$. If $z_1$ and $z_2$ are displacements of nuclei from their equilibrium positions along z-

Stable configuration of a molecule as spontaneous symmetry breaking axis and $u_1 \stackrel{def}{=} \sqrt{M_1} z_1$, $u_2 \stackrel{def}{=} \sqrt{M_2} z_2$, then the part of nuclear Hamiltonian responsible for the z-axis motions is (apart from a constant):

$$\hat{H}_{n,z} = -\frac{\hbar^2}{2}\frac{\partial^2}{\partial u_1^2} - \frac{\hbar^2}{2}\frac{\partial^2}{\partial u_2^2} + \frac{1}{2} p \cdot (u_1; u_2) \hat{W} \begin{pmatrix} u_1 \\ u_2 \end{pmatrix}, \qquad (53)$$

where $p \stackrel{def}{=} \dfrac{6 q_1 q_2}{8\pi\varepsilon_0 |\vec{\rho}_0|^3}$, $(u_1; u_2) \stackrel{def}{=} \begin{pmatrix} u_1 \\ u_2 \end{pmatrix}^T$ and $\hat{W} \stackrel{def}{=} \begin{pmatrix} 1/M_1 & -1/\sqrt{M_1 M_2} \\ -1/\sqrt{M_1 M_2} & 1/M_2 \end{pmatrix}$. The matrix $\hat{W}$ has $1/M_1 + 1/M_2$ as the only non zero eigenvalue. It follows that in the lowest order approximation for $E$ the formula for the vibration frequency of any diatomic molecule reads:

$$\omega \propto \left[ \frac{6 q_1 q_2}{8\pi\varepsilon_0 |\vec{\rho}_0|^3} \left( \frac{1}{M_1} + \frac{1}{M_2} \right) \right]^{1/2}. \qquad (54)$$

As an example for the molecule H$_2$ and the bond length of 0.07414 *nm* the vibration energy $\hbar\omega$ is around 7566 cm$^{-1}$. This is correct order of magnitude but the discrepancy with measured value is too large for formula to be of practical use. Still one has to keep in mind that the formula (54) was derived in the lowest non-trivial approximation for the electronic energy.

## 6. Conclusions
We used the following three fundamental symmetries: isotropy of space, homogeneity of space and indistinguishability of identical particles to find a group *G* of coordinate transformations which are the symmetries of the electronic effective potential. The electronic effective potential results from the Born-Oppenheimer approximation in a system of nuclei and electrons forming a molecule via Coulomb interaction. The same symmetry *G* has the potential seen by nuclei in the molecule. From the symmetry *G* we derived the formula for the dynamical representation. This formula one uses for the symmetry analysis of normal modes of a molecule and was postulated in the past on a different ground. The symmetry *G* allows considering stable configuration of a molecule from the point of view of spontaneous symmetry breaking theory. In that way we used the symmetry *G* to predict some of stationary points of the nuclear potential. We have expanded the electronic effective potential with respect to nuclear coordinates around zero, in order to predict some of the existing molecular shapes. In this expansion, the symmetry *G* that we found reduces significantly number of unknown parameters. For example in a molecule of the type $\underline{X}_4$ the symmetry *G* reduces this number from 90 to one thus underlining the power of our method. One can also use the expansion of *E* to find the relations between equilibrium distances (bond lengths) among nuclei in a molecule and its vibration frequencies. We derived one such formula for the case of any diatomic molecule.

Stable configuration of a molecule as spontaneous symmetry breaking

**Acknowledgments.** This work is supported by Serbian Ministry of Education and Science under projects OI 171005 and III 45016, and by EU under project EU FP7 NIM_NIL.